\documentclass[aps,prl,twocolumn,showpacs,amsmath,amssymb,floatfix,superscriptaddress,notitlepage]{revtex4-1}
\usepackage{color}
\usepackage{bbm}
\usepackage{graphicx}% Include figure files
\usepackage{dcolumn}% Align table columns on decimal point
\usepackage{array}
\usepackage{float}
\usepackage{supertabular}
\usepackage{longtable}
\usepackage{mathrsfs}
\usepackage{txfonts}
\usepackage{wasysym}
\usepackage[T1]{fontenc}
\usepackage{amssymb}
\usepackage[usenames,dvipsnames]{xcolor}
\usepackage{amsmath}
\usepackage{cases}
\usepackage{amstext}
\usepackage{latexsym}
\usepackage[colorlinks=true,citecolor=Cerulean,linkcolor=RubineRed,urlcolor=Cerulean]{hyperref}

\begin{document}
\title{Quantum Speed Limits 
Across the Quantum-to-Classical Transition}
%\author{dC$^3$}
\author{B. Shanahan}
\affiliation{Department of Physics, University of Massachusetts, Boston, MA 02125, USA}
\author{A. Chenu}
\affiliation{Massachusetts Institute of Technology, 77 Massachusetts Avenue, Cambridge, MA 02139, USA}
\author{N. Margolus}
\affiliation{Massachusetts Institute of Technology, 77 Massachusetts Avenue, Cambridge, MA 02139, USA}
\author{A. del Campo}
\affiliation{Department of Physics, University of Massachusetts, Boston, MA 02125, USA}

\def\L{{\rm \hat{L}}}
\def\q{{\bf q}}
\def\l{\left}
\def\r{\right}
\def\te{\mbox{e}}
\def\d{{\rm d}}
\def\t{{\rm t}}
\def\K{{\rm K}}
\def\N{{\rm N}}
\def\H{{\rm H}}
\def\la{\langle}
\def\ra{\rangle}
\def\om{\omega}
\def\Om{\Omega}
\def\vep{\varepsilon}
\def\wh{\widehat}
\def\tr{{\rm Tr}}
\def\da{\dagger}
\def\iz{\left}
\def\zi{\right}
\newcommand{\beq}{\begin{equation}}
\newcommand{\eeq}{\end{equation}}
\newcommand{\beqa}{\begin{eqnarray}}
\newcommand{\eeqa}{\end{eqnarray}}
\newcommand{\intf}{\int_{-\infty}^\infty}
\newcommand{\into}{\int_0^\infty}
\newcommand{\ket}[1]{| #1 \rangle}
\newcommand{\bra}[1]{\langle #1 |}
\newcommand{\braket}[2] {\langle #1 | #2 \rangle}
\def\bx{{ x}}
\newcommand{\bp}{{ p}}

\newcommand{\aurelia}{\color{blue}}

\newcommand{\note}{\color{green}}

\begin{abstract}
Quantum speed limits set an upper bound to the rate at which a quantum system can evolve.
Adopting a  phase-space approach we explore quantum speed limits across the quantum to classical transition and identify equivalent bounds in the classical world. As a result, and contrary to common belief, we show that speed limits exist for both   quantum and classical systems. As in the quantum domain, classical speed limits are  set by a given norm of the generator of time evolution.

\end{abstract}

\maketitle

The multi-faceted nature of time makes its treatment challenging in the quantum world  \cite{TQM1,TQM2}. Nonetheless, the understanding of time-energy uncertainty relations is somewhat privileged \cite{Busch08,Schulman08}. To a great extent, this is due to their  reformulation in terms of quantum speed limits (QSL) concerning the ability to distinguish two quantum states connected via time evolution.  While QSL provide fundamental constraints to the pace at which quantum systems can change, a plethora of applications have been found that  well extend beyond the realm of quantum dynamics. Indeed, QSL provide limits to the computational capability of physical devices \cite{Lloyd00},  the performance of quantum thermal machines in finite-time  thermodynamics \cite{delcampo14, Modi17}, parameter estimation in quantum metrology \cite{rafal,BD17},  quantum control \cite{Demirplak08,Caneva09,DRZ12,CD17,Funo17}, the decay of unstable quantum systems \cite{MT45,Bhattacharyya83,Chenu17,Beau17} and information scrambling \cite{DMS17}, among other examples \cite{Busch08,Schulman08,DC17}.

Specifically, QSL are derived as upper bounds to the rate of change of the fidelity $F(\tau)=|\la \psi_0|\psi_\tau\ra|^2\in[0,1]$ between an initial quantum state $|\psi_0\ra$ and the corresponding 
time-evolving state $|\psi_\tau\ra=\hat{U}(\tau,0)|\psi_0\ra$, where $\hat{U}(\tau,0)$ is the time-evolution operator.
More generally quantum states need not be pure, and given two density matrices $\rho_0$ and $\rho_\tau=\hat{U}(\tau,0)\rho_0\hat{U}(\tau,0)^\dag$ the fidelity reads 
\beqa
F(\tau)=\left[{\tr \sqrt{\sqrt{\rho_0}\,\rho_\tau \sqrt{\rho_0}}} \right]^2\, .
\eeqa
The fidelity is useful to define a metric between quantum states in Hilbert space, known as the  Bures angle,  \cite{Wootters81,Uhlmann92}
\begin{equation}
\label{L_def}
\mathcal{L}\left(\rho_0,\rho_\tau\right)=\cos^{-1}\left(\sqrt{F\left(\tau\right)} \right)\, .
\end{equation}
This gives a geometric interpretation of speed limit as the minimum time required to sweep out the angle $\mathcal{L}\left(\rho_0,\rho_\tau\right)$ under a given dynamics \cite{Russell17}.

For unitary processes, two seminal results are known. The Mandelstam-Tamm bound estimates the speed of evolution in terms of the energy dispersion of the initial state \cite{MT45,Fleming73,Bhattacharyya83,AA90,Vaidman92,Uhlmann92,Pfeifer93}. 
Its original derivation relies on the Heisenberg uncertainty relation. 
The second seminal result is named after Margolus and Levitin, and provides an upper bound to the speed of evolution in term of the difference between the mean energy and the ground state energy \cite{ML98,LT09}.
Its original derivation relies on the study of the survival amplitude $\la \psi_0|\psi_\tau\ra$. 
These bounds can be extended to  driven and open quantum systems \cite{QSLopen1,QSLopen2,QSLopen3,QSLopen4,QSLopen5,Pires16}.
In addition, the two bounds can be  unified \cite{LT09} so that the  time of evolution $\tau$ required to sweep an angle $\mathcal{L}\left(\rho_0,\rho_\tau\right)$ is lower bounded by 
\begin{equation}
\label{QSLbounds}
\tau\geq \tau_\mathrm{QSL}=\hbar\,\mathcal{L}\left(\rho_0,\rho_\tau\right){\rm max}\left\{\frac{1}{E-E_0},\frac{1}{\Delta E}\right\}\,,
\end{equation}
where $E_0$ is the ground state of the system, $E$ is its mean energy, and $\Delta E$  denotes the energy dispersion.  Note however that there is an infinite family of bounds in terms of higher order moments of the energy of the system \cite{ZZ06}.

It is widely believed that these bounds are quantum in nature and that, as a result, exist only in the quantum world \cite{LT09}.
Indeed, in the limit of vanishing $\hbar$, the right-hand side of (\ref{QSLbounds}) equals zero and one is led to conclude that no ``classical'' speed limit exists as the inequality becomes trivial, 
\beqa
\tau\geq \lim_{\hbar\rightarrow 0}\tau_\mathrm{QSL}= 0\, .
\eeqa
This conclusion is further supported by the aforementioned derivations of QSL, which strongly rely on the framework of quantum theory. In particular, the Mandelstam-Tamm bound follows from the Heisenberg uncertainty relation \cite{MT45,Busch08}, and the Margolus-Levitin inequality exploits the notion of the transition probability amplitude between two quantum states in Hilbert  space \cite{ML98,LT09}.
We note however that recent developments on the generalization of QSL to  open quantum systems and arbitrary quantum channels have provided  new derivations and an alternative understanding of QSL \cite{QSLopen1,QSLopen2,QSLopen3,QSLopen4,QSLopen5,Pires16}. As a result of these works, given an equation of motion for the state of the system, QSL are derived in terms of a given norm of the generator of evolution acting on the initial state of the system $\rho_0$ or the time-dependent state $\rho_t$ (with $0\leq t\leq \tau$). Such formulation appears not to be  restricted to quantum mechanical systems, as we show here.

In this Letter,  we focus on the existence and characterization of QSL across the quantum-to-classical transition.
We show that the conclusion on the quantum nature of QSL is unjustified. We demonstrate that, contrary to common belief,  similar speed limits hold in the classical world. 
To this end, we adopt a phase-space formulation of quantum mechanics and derive quantum speed limits for quasi-probability distributions;  the Wigner function. We find that the speed of evolution is determined by a certain norm of the Moyal product of the Hamiltonian and the Wigner function.
Using a semiclassical expansion, we then identify a classical speed limit   and show that the resulting bound does indeed govern the  evolution of the classical phase-space probability distribution. As a result, we establish the universal existence of fundamental limits to the pace of evolution of a physical system, independently of its classical or quantum nature.

{\it Quantum Speed Limits in phase space.---}
For simplicity and without loss of generality, we consider a one-dimensional system for which the phase-space representation is given by the Wigner function defined as  \cite{Wigner32,Hillery84}
\beqa
W_t(q,p)= 	\frac{1}{\pi\hbar} \int \left \langle  q -y\bigg| \,\hat{\rho}_t  \,\bigg| q + y  \right\rangle e^{2i p y/\hbar}  d y\, ,
\eeqa
where $\langle  q | \hat{\rho}_t  | q'  \rangle$ denotes a density matrix in the coordinate representation. 
%We assume a  one-dimensional system for clarity of presentation, the generalization to arbitrary dimension being straightforward. 
It is well known that $W_t$ is a quasi-probability distribution that takes real but possibly negative values.
We consider the Wigner function of the initial state $W_0$ and of the time-dependent state $W_t$ generated via unitary dynamics with a time-independent Hamiltonian. The fidelity between any two pure states with respective density matrices $\hat{\rho}_0$ and $\hat{\rho}_t$ can be obtained as the trace in phase space of the corresponding Wigner functions,  
\beqa
\label{fideq}
F(t)=\tr(\hat{\rho}_0\hat{\rho}_t)=\int d^2\Gamma W_0 W_t\, ,
\eeqa
where $d^2\Gamma=2\pi\hbar dqdp$, for short. 
%Note that  $\int d^2\Gamma W_0^2 = \tr(\rho_0^2) = 1$ provided the state $\rho_0$ is pure.
% which gives the Wigner functions  dimensional to $[\hbar^{-1}]$ and the fidelity  dimensionless.  

To derive a QSL, we compute the instantaneous rate of change of the fidelity as a function of time. This can be done using
the equation of motion of the Wigner function
\beqa
\label{weom}
\frac{\partial W_t}{\partial t}=\{\!\{H,W_t\}\!\}=\frac{1}{i\hbar} \left(H_{qp} \star W_t - W_t\star H_{qp} \right)\, ,
\eeqa
where the Moyal bracket $\{\!\{A,B\}\!\}$ can be explicitly written in terms of  
the Moyal product 
\beqa
	H_{qp} \star W_{t} \equiv H_{qp} \exp \! \left( 
	  \frac{i\hbar}{2} \, \overleftarrow{\partial_q} \, 
	  \overrightarrow{\partial_p} - 
	   \frac{i\hbar}{2} \, 
	   \overleftarrow{\partial_p} \, \overrightarrow{\partial_q}
	  \right)  W_{t}(q,p)\, , 
\eeqa
and where $H_{qp}=\int dx\la q-x/2|\hat{H}|q+x/2\ra \exp(ipx/\hbar)$ denotes the Weyl ordered Hamiltonian operator in phase space.
From Eqs. (\ref{fideq}) and (\ref{weom}), it follows that the rate of change of the fidelity is set by
\beqa
\dot{F}(t)&=&\int d^2\Gamma W_0\{\!\{H,W_t\}\!\}\nonumber\\
&=&\int d^2\Gamma \{\!\{H,W_0\}\!\}W_t\, , 
\eeqa
where we have used integration by parts to derive the second line. 
Using the Cauchy-Schwarz inequality one finds
\beqa
|\dot{F}(t)|\leq \left(\int d^2\Gamma W_t^2\int d^2\Gamma \{\!\{H,W_0\}\!\}^2\right)^{\frac{1}{2}}\, .
\eeqa
The purity of a density matrix is always lower than or equal to unity, so $\int d^2\Gamma W_t^2\leq 1$, where the equality is reached for pure states or  unitarity dynamics, as considered here. 
As a result, 
\beqa \label{Fdot}
|\dot{F}(t)|\leq v_{\Gamma}:=\left(\int d^2\Gamma \{\!\{H,W_0\}\!\}^2\right)^{\frac{1}{2}}\, ,
\eeqa
and we find an upper bound  $v_{\Gamma}$  to the speed of evolution in phase space,  with dimension of frequency. 
This bound is in fact dictated by the energy variance of the initial state, and  for pure states  $\nu_\Gamma = \sqrt{2}\Delta E / \hbar$, with $\Delta E  = \sqrt{\la H^2\ra - \la H\ra^2}$, as we show in \cite{SMtext}. 
A time integration between $t=0$ to $t=\tau$ readily  gives 
\begin{equation}
\frac{ 1- F(\tau)}{v_\Gamma} =     \tau_{\rm QSL}\leq \tau\, , 
\end{equation}
which is already  a QSL in phase space. 
Making use of the fact that $0\leq F(t)\leq 1$ to parameterize the fidelity in terms of the Bures angle 
\beqa
\mathcal{L}\left(\rho_0,\rho_t\right)
=\cos^{-1}\left(\sqrt{\int d^2\Gamma W_0W_t}\, \right)\, , 
\eeqa
that satisfies $F(t) = 1- \sin^2\mathcal{L}_t$,  we can rewrite
 the phase-space  QSL  as
\beqa \label{QSL}
\tau_{\rm QSL}= \frac{\sin^2\left(\mathcal{L}\left(\rho_0,\rho_\tau\right)\right)}{v_{\Gamma}}\,    = \frac{1-F(\tau) }{\sqrt{2}} \frac{\hbar}{\Delta E}.
\eeqa
Equation (\ref{QSL}) constitutes a QSL of the Mandelstam-Tamm type for the Wigner function in phase space quantum mechanics. 
The upper bound to the speed of evolution in phase space $v_{\Gamma}$ has units of frequency and is set by the action of the Moyal bracket on the initial Wigner function, that is related to the energy variance of the initial state. The distance between states is defined by the Bures angle $\mathcal{L}\left(\rho_0,\rho_t\right)$ as a natural statistical distance  \cite{Wootters81}, that is dimensionless and  independent of $\hbar$.
Note however that it is possible to derive alternative QSL by considering other distances either in the space of density operators \cite{Pires16} or in phase space \cite{Deffner17}. 
In what follows, we first use a semi-classical expansion to identify a semi-classical speed limit, and then combine the results with an operational treatment of quantum dynamics to identify a classical speed limit.

{\it Speed limits across the quantum-to-classical transition.---}
We recall that the Moyal bracket (\ref{weom}),  in a $\hbar$-expansion,  reduces to the Poisson bracket so that  
\beqa
\{\!\{W_t,H\}\!\}=\{W_t,H\}+\mathcal{O}(\hbar^2)\, , 
\label{MoyalPoisson}
\eeqa
where the action of the Poisson bracket on a function $f$ is given by
\beqa
\{f,H\}=
\frac{\partial H}{\partial p}\frac{\partial f }{\partial q}-\frac{\partial H}{\partial q}\frac{\partial  f }{\partial p}\, ,
\eeqa
and  rules the dynamics  in classical statistical mechanics according to  the (classical) Liouville equation.
As a result, to leading order in the semiclassical $\hbar$-expansion of the equation of motion for the Wigner function Eq. (\ref{weom}), the speed limit in phase space does not vanish. In particular,  the semiclassical speed limit (SSL) reads 
\beqa
\label{SSL}
\tau\geq \tau_{\rm SSL}&=&\frac{ \sin^2\mathcal{L}\left(\rho_0,\rho_\tau\right)}{\left(\int d^2\Gamma \{H,W_0\}^2\right)^{\frac{1}{2}}}\nonumber\\
&=&\frac{ \sin^2\mathcal{L}\left(\rho_0,\rho_\tau\right)}{\| \{H,W_0\} \|_2}\, ,
\eeqa
where $\| f \|_2=(\int |f|^2d\Gamma)^{1/2}$ is the $L^2$-norm of $f$ and we emphasize that $\| \{H,W_0\} \|_2$ has frequency units.

Let us discuss this expression in detail. 
The Moyal product provides a one-parameter deformation of the noncommutative algebra in quantum mechanics and of the commutative algebra in classical phase space according to Eq. (\ref{MoyalPoisson}). By reformulating QSL in terms of Wigner functions, this correspondence leads to the identification of a semiclassical speed limit (SSL) in phase space.
The distance $\mathcal{L}\left(\rho_0,\rho_\tau\right)$ between states $\rho_0$ and $\rho_\tau$ is well defined whether these states are valid classical states (i.e., with a positive Wigner function) or not. 
As a result, equation (\ref{SSL})  constitutes the semiclassical limit of  the Mandelstam-Tamm time-energy uncertainty relation.
Using Hamilton's equation of motion,  
\beqa
\frac{\partial W_t}{\partial t}=\{H,W_t\}\, , 
\eeqa
we  interpret the  upper bound to the speed of evolution as the root mean square  of the initial  rate of change of the Wigner function at $t=0$ averaged over phase space, i.e.,
\beqa
v_{\Gamma}^{\rm SSL}=  \| \{H,W_0\} \|_2 
%=\sqrt{\la (\partial_t W_t|_{t=0})^2\ra_{\Gamma}}\, .
=\sqrt{\int d^2 \Gamma (\partial_t W_t|_{t=0})^2 }\, .
\eeqa
Alternatively, introducing the Liouvillian $i\hat{L}W_t=-\{H,W_t\}$ we can restate the SSL as
\beqa
\label{qcsl}
\tau_{\rm SSL}=\frac{ \sin^2\mathcal{L}\left(W_0,W_\tau\right)}{\| \hat{L}W_0 \|_2}\, .
\eeqa
As in the quantum case (\ref{QSL}), the SSL is set by a given norm of the generator of evolution $\hat{L}$ averaged over the initial state $W_0$. 
We  note that this expression still contains an explicit $\hbar$ both in the integration measure and in the definition of the Wigner function.

{\it Classical speed limit.---}
To identify a classical speed limit (CSL) from the semiclassical expression (\ref{qcsl}), we resort to the operational dynamic modeling  developed by Bondar et al. \cite{Bondar12,Bondar13}.
The equivalence of the evolution of dynamical average
values in the quantum and classical domain via  Ehrenfest theorems yields a relation between  the classical phase-space probability density  $\varrho_t (q,p)$ and the Wigner function $W_t(q,p)$
\beqa
\label{qclcorrespondence}
\varrho_t (q,p)=2\pi\hbar W_t(q,p)^2\, .
\eeqa
Note that the factor $2\pi \hbar$, so far accounted for in $d^2\Gamma$, can be interpreted as dividing the phase-phase into cells of area $2 \pi \hbar$ \cite{Landau}, which  corresponds to the B\"ohr-Sommerfeld quantization rule in ``old'' quantum theory. The normalization of a pure quantum state  $\ket{\psi_t}$ carries over the classical distribution  $ \int 2 \pi \hbar dx dp W_t (x,p)^2 =\int dx dp \varrho_t(x,p)= 1$.

Accordingly, the fidelity (\ref{fideq}) reduces to the Bhattacharyya coefficient \cite{Bhatta46}
\beqa
{\rm B}(t)={\rm B}(\varrho_0,\varrho_t)=\int dqdp \sqrt{\varrho_0(q,p)\varrho_t(q,p)}
\eeqa
that is related to the Hellinger distance ${\rm H}(\varrho_0,\varrho_t)$  via the identity ${\rm B}(t)=1-{\rm H}(\varrho_0,\varrho_t)^2$.
Note that ${\rm B}(0)=1$ due to the normalization condition. % $\int dqdp\varrho_0=1$. 
The Bures angle becomes 
\beqa
\mathcal{L}_{\rm B}=\cos^{-1}\sqrt{{\rm B}(t)}\, ,
\eeqa
and the classical speed limit (CSL) thus reads $\tau\geq \tau_{\rm CSL}$ with 
\beqa
\label{csl}
\tau_{\rm CSL}
&=&\frac{ \sin^2\mathcal{L}_{\rm B}\left(\varrho_0,\varrho_\tau\right)}{\sqrt{\int dqdp (\partial_t\sqrt{\varrho_t}|_{t=0})^2}}=\frac{\sin^2\mathcal{L}_{\rm B}\left(\varrho_0,\varrho_\tau\right)}{\sqrt{\int dqdp \left\{H,\sqrt{\varrho_0}\,\right\}^2}}\nonumber\\
&=&\frac{  1 - B(\tau) }{\| \hat{L}\sqrt{\varrho_0}\|_2}\, ,
\eeqa
where $\hat{L}$ is the classical Liouville operator satisfying $\partial\varrho_t+i\hat{L}\varrho_t=0$.
This is our main result and constitutes a classical version of the Mandelstam-Tamm bound.

It is worth emphasizing that this bound can be derived independently of the semiclassical approach by making exclusive reference to the classical Hamiltonian formalism.
Indeed, the rate of change of the  Bhattacharyya coefficient is given by
\beqa
\dot{{\rm B}}(\varrho_0,\varrho_t)=\int dqdp\sqrt{\rho_0}\frac{\dot{\varrho}_t}{2\sqrt{\varrho_t}}\, .
\eeqa
Using Liouville's equation, we can rewrite the rate of change of the classical probability distribution to find
\beqa
\frac{\dot{\varrho}_t}{2\sqrt{\varrho_t}}=\frac{\{H,\varrho_t\}}{2\sqrt{\varrho_t}}=\left\{H,\sqrt{\varrho_t}\, \right\}\, .
\eeqa
To obtain a classical speed limit that depends only on the initial state, as opposed to its time evolution, it is convenient to shift the action of the Poisson bracket to the initial state $\varrho_0$.
This is readily accomplished by integration by parts, assuming $\varrho_t$ vanishes at the end points of integration, that  yields
\beqa
\label{Bdot}
\dot{{\rm B}}(\varrho_0,\varrho_t)&=&-\int dqdp\left\{H,\sqrt{\rho_0}\,\right\}\sqrt{\varrho_t}\, .
\eeqa
Use of the Cauchy-Schwarz inequality  and the normalization condition $ \int dqdp\varrho_t=1$ lead to
\beqa
|\dot{{\rm B}}(\varrho_0,\varrho_t)|\leq\left(\int dqdp\left\{H,\sqrt{\rho_0}\,\right\}^2\right)^{\frac{1}{2}}\, ,
\eeqa 
which upon integration over the time variable from $t=0$ to $t=\tau$ yields Eq (\ref{csl}), given that $1-{\rm B}(t)=\sin^2\mathcal{L}_{\rm B}$.
Note that we consider only smooth classical phase-space distributions, for which $v_{\Gamma}^{\rm CSL}=\|\partial_t\sqrt{\varrho_t}|_{t=0}\|_2$ is well-defined.  For a singular distribution of the form $\varrho_t(q,p)=\delta[q-q_{\rm cl}(t)] \delta[p-p_{\rm cl}(t)]$, characterizing a certain trajectory of a classical particle, the upper bound to the phase-space velocity $\| \hat{L}\sqrt{\varrho_0}\|_2$ is singular and needs to be regularized. In this limit,  the CSL is expected to vanish as the the trajectories  $\varrho_t(q,p)$ and $\varrho_t(q,p)'=\varrho_t(q+\epsilon_q,p+\epsilon_p)$ are distinguishable for any $\epsilon_q$, $\epsilon_p$ with $|\epsilon_q|>0$ and $|\epsilon_p|>0$ in the sense that ${\rm B}(\varrho_0,\varrho_t')=0$ and $\mathcal{L}_{\rm B}=\pi/2$. 

{\it Quadratic Hamiltonians.---} The existence of classical speed limits and their correspondence with their quantum counterpart  become self-evident whenever the Hamiltonian driving the evolution is quadratic in the position and momentum operators. The equation of motion of the Wigner function (\ref{weom}) simplifies and the phase-space generators of evolution in  classical  and quantum dynamics  are then equivalent. In the classical case, for a time-independent Hamiltonian the corresponding canonical transformations, 
\beqa
\begin{pmatrix} 
q \\
p
\end{pmatrix}
=
\begin{pmatrix}
\alpha & \beta \\
\gamma & \delta
\end{pmatrix}
\begin{pmatrix}
q' \\
p'
\end{pmatrix}\, ,
\eeqa 
are elements of the two-dimensional real symplectic group $Sp(2,\mathbb{R})$. In the quantum case, the phase-space propagator that determines the evolution of the Wigner function via the identity
\beqa
W_n(q,p;t)=\iint dq'dp'K(q,p|q',p')W_n(q',p';0)
\eeqa
becomes
\beqa \label{K_HO}
K(q,p|q',p')=\delta[q'-(\alpha q+\beta p)]\delta[p'-(\gamma q+\delta p)]\, ,
\eeqa
and it is therefore identical to the classical one \cite{GCM90}. The quantum and semiclassical phase-space limits, Eqs. (\ref{QSL}) and (\ref{SSL}), are identical in this case.
When the generator  of evolution is explicitly time-dependent, a representation of the corresponding canonical transformations is still possible. For the sake of illustration we focus on the time-dependent harmonic oscillator,
\beqa
\label{Hamho}
\hat{H}=\frac{\hat{p}^2}{2m}+\frac{1}{2}m\om(t)^2\hat{q}^2\, ,
\eeqa 
 for which quantum speed limits have been reported with multiple applications including the characterization of control protocols \cite{CM10,CCM16,Zheng16,CD17,Funo17} and the performance of quantum thermal machines \cite{delcampo14}. 
As shown in \cite{SMtext}, in the quantum case, the Wigner function of an eigenstate at $t=0$ evolves under a modulation of the trapping frequency $\om(t)$ according to 
\beqa
& & W_n(q,p;t)=W_n\left(\frac{q}{b},bp-mq\dot{b};0\right)\\
&=&\frac{(-1)^n}{\pi\hbar}e^{-\frac{2}{\hbar\om_0}\left(\frac{P^2}{2m}+\frac{1}{2}m\om_0^2Q^2\right)}
%\nonumber\\
%& & \times 
L_n\left[\frac{4}{\hbar\om_0}\left(\frac{P^2}{2m}+\frac{1}{2}m\om_0^2Q^2\right)\right]\, ,\nonumber
\eeqa
that we explicitly find in terms of the Laguerre polynomials $L_n(x)$ and the canonically conjugated pair of variables
\beqa
\label{conjpair}
Q:=\frac{q}{b},\quad P=bp-mq\dot{b}
\eeqa
associated with the matrix
$
\begin{pmatrix}
\alpha & \beta \\
\gamma & \delta
\end{pmatrix}=
\begin{pmatrix}
1/b & 0 \\
-m\dot{b} & b
\end{pmatrix}\, .
$
The time-dependent scaling factor $b(t)>0$ is the solution of the Ermakov equation, $\ddot{b}+\om(t)^2b=\om_0^2/b^3$, with the boundary conditions $b(0)=1$ and $\dot{b}(0)=0$; see e.g. \cite{Chen10}.
As a result, the dynamics arbitrarily far from equilibrium does not alter the form of the Wigner function and can be simply accounted for by the definition of the conjugated pair (\ref{conjpair}). 

\begin{figure}
 \includegraphics[width= 0.7\columnwidth]{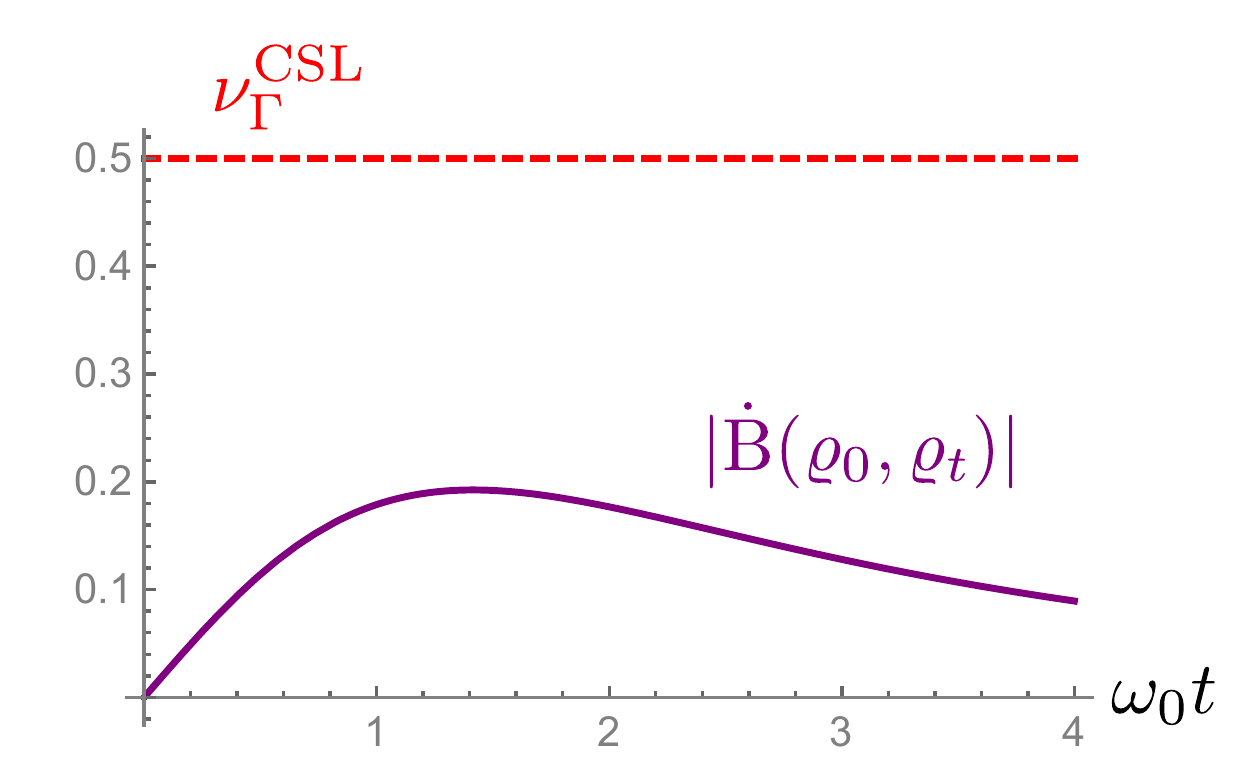}\ \ \ \\
\caption{{\bf Classical speed limit to the pace of evolution.}  Comparison of the upper bound to the phase-space speed of evolution $v_{\Gamma}^{\rm CSL}$ with the absolute value of the instantaneous rate of change of the Battacharyya coefficient $|\dot{{\rm B}}(\varrho_0,\varrho_t)|$ as  a function of time. The dynamics corresponds to a free expansion of a classical probability distribution of Gaussian form that is initially confined in a harmonic potential of frequency $\om_0$, which is switched off for $t>0$. The unit of time is set by  $\om_0^{-1}$.}\label{Fig1}
\end{figure}

For the ground-state of the harmonic oscillator with $n=0$, $W_0(q,p,t)\geq 0$ is a smooth Gaussian distribution for all $0\leq t\leq \tau$. 
When the  classical  distribution is chosen to be also of Gaussian form $\rho_0(q,p)=\exp(-q^2/\sigma_q^2-p^2/\sigma_p^2)/(\pi\sigma_q\sigma_p)$ the CSL in Eq. (\ref{csl})
equals the  quantum and semiclassical phase-space limits, Eqs. (\ref{QSL}) and (\ref{SSL}), provided that $\sigma_q=x_0/\sqrt{2}$ and $\sigma_p=\hbar/(x_0\sqrt{2})$ as dictated by the correspondence (\ref{qclcorrespondence}); see \cite{SMtext}.
From the exact dynamics  $\rho_t(q,p)=\rho_0(Q,P)$, we find the Bhattacharyya coefficient
\beqa
{\rm B}(\varrho_0,\varrho_t)=2\left[\frac{(1+b^2)^2}{b^2}+\left(\frac{m\sigma_x\dot{b}}{\sigma_p}\right)^2\right]^{-\frac{1}{2}}\, ,
\eeqa
while  the upper bound for the phase-space speed of evolution is set by
\beqa
\label{tdhov}
v_{\Gamma}^{\rm CSL}=\left\|\left\{H,\sqrt{\rho_0}\,\right\}\right\|_{\rm 2}=\frac{m\sigma_x|\ddot{b}(0)|}{2\sigma_p}\, .
%\sqrt{\frac{\ddot{b}(0)^2}{\om_0^2}}.
\eeqa
 While the generalization of the CSL (\ref{csl}) to time-dependent generators is straightforward \cite{SMtext} we focus on the case when the driven Hamiltonian is constant for $t>0$ and let the frequency of the trap  be suddenly turned off at $t=0$. It then follows that $b(t)=\sqrt{1+\om_0^2t^2}$ and  $\ddot{b}(0)=\om_0$. To illustrate these results, we show in Figure \ref{Fig1}  how the characteristic velocity in phase space of $v_{\Gamma}^{\rm CSL}$ in (\ref{tdhov}) remains an upper bound to  the instantaneous phase-space velocity set by the absolute value of the Bhattacharyya coefficient during the course of the evolution.

In conclusion, we have shown that there exist fundamental speed limits to the pace of evolution of an arbitrary physical system, both in the classical and quantum worlds. 
To this end, we have introduced quantum speed limits in phase space and derived their semiclassical limit. Their comparison should be useful to  identify  scenarios in which the quantum dynamics provides a speedup over   the classical evolution. From the semiclassical limit, we have further identified a  family of classical speed limits that governs the  classical Hamiltonian dynamics in phase space. 
In the quantum, semiclassical and classical settings, speed limits are universally set  by a given norm of the generator of the dynamics and the state of the system under consideration. 
Our results provide further insight on the nature of time-energy uncertainty relations, speed limits in arbitrary physical process and onto the limits of computation.

{\it Note.---} After  the  completion  of  this  work,  we  learned about reference  \cite{OO17} devoted to  classical speed limits in Hilbert space.

{\it Acknowledgments.---} It is a great pleasure to acknowledge discussions with I. L. Egusquiza and L. P. Garc\'ia-Pintos.
Funding support from the John Templeton Foundation, UMass Boston (project P20150000029279), and the National Institute of General Medical Sciences of the National Institutes of Health (Award Number R25GM076321) is greatly acknowledged.
AC and ADC gratefully acknowledge support from the Simons Center for Geometry and Physics, Stony Brook University, during the completion of this project.

\newpage
\begin{widetext}

\appendix

\section{Supplementary Material}

\subsection{Operational interpretation of  the phase space speed of evolution $\nu_\Gamma$}
We show below that the frequency $\nu_{\Gamma}$, that sets an upper bound to the speed of evolution in phase-space quantum mechanics, can be related to the energy variance. Starting with the definition given in Eq. (11) in the main text, we have 
\begin{equation}
\nu_\Gamma^2 =\int d^2\Gamma \{\!\{H,W_0\}\!\}^2 = \int d^2\Gamma\left(\left. \frac{\partial W_t}{\partial t} \right \vert_{t=0}\right)^2.
\end{equation}
From the definition of the Wigner function, Eq. (5), the time derivative can be written as 
\begin{equation}
\frac{\partial W_t}{\partial t} = 	\frac{1}{\pi\hbar} \int \left \langle  q -y\bigg| \, \frac{\partial \hat{\rho}_t}{\partial t}  \,\bigg| q + y  \right\rangle e^{2i p y/\hbar}  d y\, ,
\end{equation}
which, for a Hermitian Hamiltonian $H = H^\dagger$ and using the fact that $\int dp \, e^{2 i p (y+y')/\hbar} =  \pi \hbar  \delta(y+y')$,  is set by the square  of the rate of change of the density matrix integrated over phase-space   
\begin{eqnarray}
 \int d^2\Gamma\left(\frac{\partial W_t}{\partial t} \right)^2 
 &=& 2 \int dqdy \bra{q-y} \frac{\partial \hat{\rho}_t}{\partial t} \ket{q+y} \bra{q+y}\frac{\partial \hat{\rho}_t}{\partial t} \ket{q-y} \\
 &=& 2 \int \frac{dX dY}{2} \bra{X} \frac{\partial \hat{\rho}_t}{\partial t} \ket{Y} \bra{Y}\frac{\partial \hat{\rho}_t}{\partial t} \ket{X} \\
 &=& \tr\left( \left(  \frac{\partial \hat{\rho}_t}{\partial t}\right)^2 \right).
\end{eqnarray}
%In the last line, we have used the closure relation $\int dx \ket{x+y}\bra{x+y} = \mathbb{1}$ to simplify the expression. 
We can further use the Heisenberg equation, $ i \hbar \frac{\partial \hat{\rho}_t}{\partial t} = [ H,\rho_t]$, to write the trace as 
\begin{equation}
\nu_\Gamma^2 = \frac{2}{\hbar^2} \tr\left( H\rho_0 \rho_0 H - (H \rho_0)^2 \right).
\end{equation}
Since for a normalized pure state $\ket{\psi}$, the density matrix is idempotent ($\rho^2 = \ket{\psi}\braket{\psi}{\psi} \bra{\psi} = \rho$), the above expression simplifies 
to $\nu_\Gamma^2 = 2 (\Delta E /  \hbar)^2$, as given in the main text.

\subsection{Computation of the fidelity for the time-dependent harmonic oscillator} 

Consider a driven harmonic oscillator with an arbitrary frequency modulation $\om(t)$ for $t\geq 0$.
In the quantum case, it is well-known that an eigenstate at $t=0$ evolves under a modulation of the trapping frequency $\om(t)$ according to a self-similar dynamics
\beqa
\label{scansatz}
\psi_n(q,t)=b^{-1/2}\exp\left(i\frac{m\dot{b}}{2\hbar b}q^2\right)\psi_n\left(\frac{q}{b},t=0\right)\, ,
\eeqa
where the time-dependent scaling factor $b(t)>0$ is the solution of the Ermakov equation, $\ddot{b}+\om(t)^2b=\om_0^2/b^3$, with the boundary conditions $b(0)=1$ and $\dot{b}(0)=0$; see e.g. [46].
The corresponding Wigner function is given by
\beqa
W_n(q,p;t)=W_n\left(\frac{q}{b},bp-mq\dot{b};0\right)=\frac{(-1)^n}{\pi\hbar}e^{-\frac{2}{\hbar\om_0}\left(\frac{P^2}{2m}+\frac{1}{2}m\om_0^2Q^2\right)}
L_n\left[\frac{4}{\hbar\om_0}\left(\frac{P^2}{2m}+\frac{1}{2}m\om_0^2Q^2\right)\right]\, ,
\eeqa
in terms of the Laguerre polynomials $L_n(x)$ and the canonically conjugated pair of variables $Q:=\frac{q}{b}$ and $ P=bp-mq\dot{b}$. 

The explicit form of the fidelity $F_n(t):=F[|\psi_n\ra\la\psi_n|, \hat{U}(t,0)|\psi_n\ra\la\psi_n|\hat{U}(t,0)^\dag]$ between an initial eigenstate of the harmonic oscillator $|\psi_n\ra$ and its time evolution $ \hat{U}(t,0)|\psi_n\ra$ can be efficiently computed in phase space noting that
\beqa
F_n(t)=\int d^2\Gamma W_n(q,p;0)W_n(q,p;t)\, .
\eeqa

For the time-dependent harmonic oscillator, the explicit integral representation of the fidelity
\beqa
F_n(t)=\frac{2}{\pi\hbar}\int dqdp \exp \left[- \frac{2}{\hbar \omega_0} \bigg( h(q, p) + h(Q, P) \bigg) \right]
L_n \left[ \frac{4}{\hbar \omega_0} h(q, p) \right] L_n \left[ \frac{4}{\hbar \omega_0} h(Q, P) \right]\, ,
\eeqa
where $ h(q, p) = \frac{p^2}{2m} + \frac{1}{2} m \omega^2 q^2$, can be conveniently rewritten in terms of the  
 generating function $g(x, z)$ of the Laguerre polynomials 
\beqa
g(x, u) = \frac{1}{1-u} \hspace{1mm} \exp \left[ - \frac{xu}{1-u} \right] = \sum_{n = 0}^{\infty} u^n L_n(x)\, ,
\eeqa
given the identity $L_n(x) = \frac{1}{n!} \frac{d^n}{du^n}g(x, u) \big|_{u = 0}$.
Therefore,
\beqa
\label{Fneq}
F_n(t)=\frac{1}{(n!)^2} \frac{d^n}{du^n}\frac{d^n}{dv^n}\mathcal{I}(u,v;t)\, ,
\eeqa
where the Gaussian integral
\beqa
\mathcal{I}(u,v;t):=\frac{2}{\pi\hbar}\int dqdp \exp \left[- \frac{2}{\hbar \omega_0} \bigg( h(q, p) + h(Q, P) \bigg) \right]
g\left( \frac{4}{\hbar \omega_0} h(q, p) , u\right) g\left( \frac{4}{\hbar \omega_0} h(Q, P) , v\right)
\eeqa
can be explicitly  found
\beqa
\label{Iuvteq}
\mathcal{I}(u,v;t)=
\frac{2 b \omega _0}{(u-1) (v-1) \sqrt{-\frac{b^2 (u-1) (v+1)+(u+1) (v-1)}{(u-1) (v-1)}} \sqrt{-\frac{b^2 \dot{b}^2 \left(u^2-1\right) \left(v^2-1\right)+\omega _0^2 \left(b^2 (u+1) (v-1)+(u-1) (v+1)\right)
   \left(b^2 (u-1) (v+1)+(u+1) (v-1)\right)}{(u-1) (v-1) \left(b^2 (u-1) (v+1)+(u+1) (v-1)\right)}}}\, .
\eeqa
Using (\ref{Fneq}) and (\ref{Iuvteq}) one can derive explicit expressions for arbitrary an quantum number $n$, e.g.,
\beqa
F_0(t)&=&\frac{2 b \omega _0}{\left[\left(b^2+1\right)^2 \omega _0^2+b^2 \dot{b}^2\right]^{1/2}}\, ,\\
F_1(t)&=&\frac{8 b^3 \omega _0^3}{\left[\left(b^2+1\right)^2 \omega _0^2+b^2 \dot{b}^2\right]^{3/2}}\, ,\\
F_2(t)&=&\frac{b \omega _0 \left[b^2 \left(\left(b^2-10\right) \omega _0^2+\dot{b}^2\right)+\omega _0^2\right]^2}{2\left[\left(b^2+1\right)^2 \omega _0^2+b^2 \dot{b}^2\right]^{5/2}}\, ,\\
F_3(t)&=&\frac{2 b^3 \omega _0^3 \left[3 b^2 \dot{b}^2+\left(3 b^4-14 b^2+3\right) \omega _0^2\right]^2}{\left[\left(b^2+1\right)^2 \omega _0^2+b^2 \dot{b}^2\right]^{7/2}}\, .
\eeqa

For $n=0$, the  pure quantum states $\hat{\rho}_0=|\psi(0)\ra\la \psi(0)|$ and $\hat{\rho}_t|\psi(t)\ra\la \psi(t)|$ have positive Wigner functions $W_0$ and $W_t$, respectively. Using the identification $\varrho_s (q,p)=2\pi\hbar W_s(q,p)^2$  for $s=0,t$ it follows that the fidelity $F(t)=\int d^2\Gamma W_0 W_t$  equals the Bhattacharyya coefficient
${\rm B}(t)$, as
\beqa
F(t)&:=&\int d^2\Gamma W_0 W_t\\
&=&\int dqdp \sqrt{\varrho_0(q,p)\varrho_t(q,p)}\\
&=:&{\rm B}(t).
\eeqa
Therefore, the existence of QSL for $F(t)$ implies the existence of CSL for ${\rm B}(t)$.

\subsection{Classical speed limits for time-dependent generators}
Bounds to the speed of evolution can be found for time-dependent generators. For quantum speed limits, this is a common approach in the literature, despite the fact that the computation of the bound requires knowledge of the exact evolution $\rho_t$. Such bounds remain useful when  expressions in closed form can be derived including all parameters of the model.
Analogously, for time-dependent generators, a classical speed limit that refers only to the initial state is expected to be poor and and alternative bound can be derived at the cost of making reference to the exact evolution of the system $\varrho_t$. 

We consider the Liouville equation with a time-dependent Hamiltonian $H(t)$. The derivation of a classical speed limit in this case is analogous to that for time-independent generators and it is actually simpler.
The rate of change of the Bhattacharyya coefficient is given by
\beqa
\dot{{\rm B}}(t)
&=&\int dqdp\sqrt{\varrho_0}\frac{\dot{\varrho}_t}{2\sqrt{\varrho_t}}\\
&=&\int dqdp\sqrt{\varrho_0}\left\{H(t),\sqrt{\varrho_t}\,\right\}, \label{b2}
\eeqa
and via the Cauchy-Schwarz inequality it follows that
\beqa
|\dot{{\rm B}}(t)|&\leq&\left[\int dqdp\left\{H(t),\sqrt{\varrho_t}\,\right\}^2\right]^{\frac{1}{2}}\\
&=:&\|\hat{L}\sqrt{\varrho_t}\|_2\, .
\eeqa
Integrating from time $t=0$ to $t=\tau$ we find
\beqa
1-{\rm B}(\tau)=\sin^2\mathcal{L}_{\rm B}(\tau)
&\leq&\int_0^\tau dt \|\hat{L}\sqrt{\varrho_t}\|_2\\
&=&\tau\overline{\|\hat{L}\sqrt{\varrho_t}\|_2}\, ,
\eeqa
where we have introduced the time-averaged phase-space classical speed bound, with
\beqa
\overline{A}=\frac{1}{\tau}\int_0^\tau dtA(t)\, .
\eeqa
As a result,
\beqa
\tau_{\rm CSL}\geq\frac{\sin^2\mathcal{L}_{\rm B}(\tau)}{\overline{\|\hat{L}\sqrt{\varrho_t}\|_2}}\, .
\eeqa

\end{widetext}


\begin{thebibliography}{99}




\bibitem{TQM1}
J. G. Muga, R. Mayato, I. L. Egusquiza.  (Eds.), {\it Time in Quantum Mechanics - Vol 1}, Lect. Notes Phys. {\bf 734} (Springer, Heidelberg, 2002).
\bibitem{TQM2}
J. G. Muga, A. Ruschaupt, A. del Campo (Eds.), {\it Time in Quantum Mechanics - Vol 2}, Lect. Notes Phys. {\bf 789} (Springer, Heidelberg, 2009).

\bibitem{Busch08} P. Busch, \href{http://dx.doi.org/10.1007/978-3-540-73473-4_3}{Lect. Notes Phys. {\bf 734}, 73 (2008)};  Chapter 3 in \cite{TQM1}.
\bibitem{Schulman08} L. S. Schulman,
% Jump time and passage time: the duration of a quantum transition,
\href{http://dx.doi.org/10.1007/978-3-540-73473-4_4}{Lect. Notes Phys. {\bf 734}, 107 (2008)};  Chapter 4 in \cite{TQM1}.


\bibitem{Lloyd00}
S. Lloyd, \href{http://dx.doi.org/10.1038/35023282}{Nature {\bf 406}, 1047 (2000)}; S. Lloyd, \href{http://dx.doi.org/10.1103/PhysRevLett.88.237901}{Phys. Rev. Lett. {\bf 88}, 237901 (2002)}; V. Giovannetti, S. Lloyd, and L. Maccone, \href{http://dx.doi.org/10.1103/PhysRevA.67.052109}{Phys. Rev. A {\bf 67}, 052109  (2003).}



\bibitem{delcampo14} A. del Campo, J. Goold, M. Paternostro, \href{http://dx.doi.org/10.1038/srep06208}{Sci. Rep. {\bf 4}, 6208 (2014).}

\bibitem{Modi17}
F. Campaioli, F. A. Pollock, F. C. Binder, L. C. C\'eleri, J. Goold, S. Vinjanampathy, K. Modi, \href{http://dx.doi.org/10.1103/PhysRevLett.118.150601}{
Phys. Rev. Lett. {\bf 118}, 150601 (2017). }

\bibitem{rafal}
R. Demkowicz-Dobrzanski, J. Kolodynski, and M. Guta, \href{http://dx.doi.org/10.1038/ncomms2067}{Nat. Commun. {\bf 3}, 1063 (2012).}


\bibitem{BD17} M. Beau and A. del Campo,  \href{http://dx.doi.org/10.1103/PhysRevLett.119.010403}{Phys. Rev. Lett. {\bf 119}, 010403 (2017) .}

\bibitem{Demirplak08} M. Demirplak and S. A. Rice, 
%{\it Adiabatic Population Transfer with Control Fields. }
\href{http://dx.doi.org/10.1063/1.2992152}{{ J. Chem. Phys.} {\bf 129}, 54111 (2008).} 

\bibitem{Caneva09}
T. Caneva, M. Murphy, T. Calarco, R. Fazio, S. Montangero, V. Giovannetti, and G. E. Santoro, \href{http://dx.doi.org/10.1103/PhysRevLett.103.240501}{Phys. Rev. Lett. {\bf 103}, 240501 (2009).}

\bibitem{DRZ12} A. del Campo, M. M. Rams, W. H. Zurek, \href{http://dx.doi.org/10.1103/PhysRevLett.109.115703}{Phys. Rev. Lett. {\bf 109}, 115703 (2012).}

\bibitem{CD17} S. Campbell and S. Deffner,  \href{http://dx.doi.org/10.1103/PhysRevLett.118.100601}{Phys. Rev. Lett. {\bf118}, 100601 (2017).}

\bibitem{Funo17} K. Funo, J.-N. Zhang, C. Chatou, K. Kim, M. Ueda, and A. del Campo, \href{http://dx.doi.org/10.1103/PhysRevLett.118.100602}{Phys. Rev. Lett. {\bf 118}, 100602 (2017).}



\bibitem{MT45}
L. Mandelstam and I. Tamm, J. Phys. (USSR) {\bf 9}, 249 (1945).

\bibitem{Bhattacharyya83}
K. Bhattacharyya, \href{http://dx.doi.org/10.1088/0305-4470/16/13/021}{J. Phys. A: Math. Gen. {\bf 16}, 2993 (1983).}

\bibitem{Chenu17}  A. Chenu, M. Beau, J. Cao, A. del Campo, \href{10.1103/PhysRevLett.118.140403}{Phys. Rev. Lett. {\bf 118}, 140403 (2017).}
\bibitem{Beau17} M. Beau, J. Kiukas, I. L. Egusquiza, A. del Campo, \href{10.1103/PhysRevLett.119.130401
}{Phys. Rev. Lett. {\bf 119}, 130401 (2017).}

\bibitem{DMS17}
A. del Campo, J. Molina-Vilaplana, J. Sonner, \href{http://dx.doi.org/10.1103/PhysRevD.95.126008}{Phys. Rev. D {\bf 95}, 126008 (2017).}

\bibitem{DC17} S. Deffner and S. Campbell, \href{https://arxiv.org/abs/1705.08023}{arXiv:1705.08023 (2017).}


\bibitem{Fleming73}
G. N. Fleming, \href{http://dx.doi.org/10.1007/BF02819419}{Nuov. Cim. {\bf 16 A}, 232 (1973).}


\bibitem{AA90}
J. Anandan and Y. Aharonov, \href{http://dx.doi.org/10.1103/PhysRevLett.65.1697}{Phys. Rev. Lett. {\bf 65}, 1697 (1990).}

\bibitem{Vaidman92}
L. Vaidman, \href{http://dx.doi.org/10.1119/1.16940}{Am. J. Phys. {\bf 60}, 182 (1992).}

\bibitem{Wootters81} W. K. Wootters, \href{https://doi.org/10.1103/PhysRevD.23.357}{Phys. Rev. D {\bf 23}, 357 (1981).}


\bibitem{Uhlmann92}
A. Uhlmann, \href{http://dx.doi.org/10.1016/0375-9601(92)90555-Z}{Phys. Lett. A {\bf 161}, 329 (1992).}

\bibitem{Russell17} B. Russell and  S. Stepney,  \href{https://doi.org/10.1142/S0129054117500204}{Int. J. Found. Comput. Sci. 28, 321 (2017).}


\bibitem{Pfeifer93}
P. Pfeifer, \href{http://dx.doi.org/10.1103/PhysRevLett.70.3365}{Phys. Rev. Lett. {\bf70}, 3365 (1993).}

\bibitem{ML98}
N. Margolus and L. B. Levitin, \href{http://dx.doi.org/10.1016/S0167-2789(98)00054-2}{Physica D {\bf 120}, 188 (1998).}

\bibitem{LT09}
L. B. Levitin and T. Toffoli, \href{http://dx.doi.org/10.1103/PhysRevLett.103.160502}{Phys. Rev. Lett. {\bf 103}, 160502 (2009).}



\bibitem{QSLopen1}
M. M. Taddei, B. M. Escher, L. Davidovich, and R. L. de Matos Filho, \href{http://dx.doi.org/10.1103/PhysRevLett.110.050402}{Phys. Rev. Lett. {\bf 110}, 050402 (2013).}

\bibitem{QSLopen2}
A. del Campo, I. L. Egusquiza, M. B. Plenio, and S. F. Huelga, \href{http://dx.doi.org/10.1103/PhysRevLett.110.050403}{Phys. Rev. Lett. {\bf 110}, 050403 (2013).}

\bibitem{QSLopen3}
S. Deffner, and E. Lutz, \href{http://dx.doi.org/10.1103/PhysRevLett.111.010402}{Phys. Rev. Lett. {\bf 111}, 010402 (2013).}

\bibitem{QSLopen4}
Y.-J. Zhang, W. Han, Y.-J. Xia, J.-P. Cao, and H. Fan, \href{http://dx.doi.org/10.1038/srep04890}{Sci. Rep. {\bf 4}, 4890 (2014).}

\bibitem{QSLopen5}
I. Marvian and D. A. Lidar, \href{https://doi.org/10.1103/PhysRevLett.115.210402}{Phys. Rev. Lett. {\bf 115}, 210402 (2015).}

\bibitem{Pires16} D. P. Pires, M. Cianciaruso, L. C. C\'eleri, G. Adesso, and D. O. Soares-Pinto,  \href{https://doi.org/10.1103/PhysRevX.6.021031}{Phys. Rev. X {\bf 6}, 021031 (2016).}

\bibitem{ZZ06} B. Zieli\'nski and M. Zych,  \href{https://doi.org/10.1103/PhysRevA.74.034301}{Phys. Rev. A {\bf 74}, 034301(2006).}



\bibitem{Wigner32} E. P. Wigner, \href{http://dx.doi.org/10.1103/PhysRev.40.749.}{Phys. Rev. {\bf 40}, 749 (1932).}

\bibitem{Hillery84} M. Hillery, R. F. O'Connell, M. O. Scully, E. P. Wigner, \href{http://dx.doi.org/10.1016/0370-1573(84)90160-1}{Phys. Rep. {\bf 106}, 121 (1984).}

\bibitem{SMtext} See Supplemental Material.

\bibitem{Deffner17} S. Deffner, \href{https://arxiv.org/abs/1704.03357}{arXiv:1704.03357 (2017).}


\bibitem{Bondar12} 
D. I. Bondar, R. Cabrera, R. R. Lompay, M. Y. Ivanov, H. A. Rabitz,  \href{https://doi.org/10.1103/PhysRevLett.109.190403}{Phys. Rev. Lett. {\bf 109}, 190403 (2012).}


\bibitem{Bondar13} D. I. Bondar, R. Cabrera, D. V. Zhdanov, H. A. Rabitz, \href{https://doi.org/10.1103/PhysRevA.88.052108}{Phys. Rev. A {\bf 88}, 052108 (2013).}

\bibitem{Landau} L.D. Landau and E.M. Lifshitz, Quantum Mechanics (Vol. 3, Second edition), Chap VII,  Pergamon Press (1965).

\bibitem{Bhatta46} A. Bhattacharyya, \href{http://www.jstor.org/stable/25047882 }{The Indian Journal of Statistics {\bf 7},  401 (1946).}

%\bibitem{Margolus16} N. Margolus,  \href{http://lanl.arxiv.org/abs/1109.4994}{arXiv:1109.4994 (2016)}.




\bibitem{GCM90}  G. Garc\'ia-Calder\'on and M. Moshinsky, \href{http://dx.doi.org/10.1088/0305-4470/13/6/004}{J. Phys. A: Math. Gen. {\bf 13},  L185 (1990).}

\bibitem{CM10} Xi Chen, J. G. Muga,  \href{https://doi.org/10.1103/PhysRevA.82.053403}{Phys. Rev. A {\bf 82}, 053403 (2010).}
\bibitem{CCM16}
Yang-Yang Cui, Xi Chen, and J. G. Muga, \href{https://doi.org/10.1021/acs.jpca.5b06090}{J. Phys. Chem. A {\bf 120}, 2962 (2016).}
\bibitem{Zheng16} Y. Zheng, S. Campbell, G. De Chiara, D. Poletti, \href{http://dx.doi.org/10.1103/PhysRevA.94.042132}{Phys. Rev. A {\bf 94}, 042132 (2016).}

\bibitem{Chen10} X. Chen, A. Ruschhaupt, S. Schmidt, A. del Campo, D. Gu\'ery-Odelin, J. G. Muga, \href{https://doi.org/10.1103/PhysRevLett.104.063002}{Phys. Rev. Lett. {\bf 104}, 063002 (2010).}

\bibitem{OO17} M. Okuyama and M. Ohzeki, \href{https://arxiv.org/abs/1710.03498}{arXiv:1710.03498 (2017)}.


\end{thebibliography}
\end{document}